
%

\input harvmac.tex      
\input epsf.tex

%
%
%
\def\infinity{\infty}
\def\bra#1{{\langle #1 |  }}

\def\bar{\overline}
\def\hat{\widehat}
\def\*{\star}
\def\[{\left[}
\def\]{\right]}
\def\({\left(}		
\def\){\right)}

%
%

\def\frac#1#2{{#1 \over #2}}
\def\inv#1{{1 \over #1}}

\def\d{\partial}

\def\vev#1{\langle #1 \rangle}
\def\ket#1{ | #1 \rangle}
\def\bra#1{\langle #1 |}

\def\2pi{\hbox{$2\pi i$}}

\def\dsl{\raise.15ex\hbox{/}\kern-.57em\partial}
\def\Dsl{\,\raise.15ex\hbox{/}\mkern-.13.5mu D}
%
%
\def\th{\theta}		
		
\def\be{\beta}
\def\al{\alpha}
\def\ep{\epsilon}

\def\om{\omega}		
\def\sig{\sigma}	

%
%
\def\CA{{\cal A}}		
	\def\CE{{\cal E}}

\def\2pi{\hbox{$2\pi i$}}

\def\dsl{\raise.15ex\hbox{/}\kern-.57em\partial}
\def\Dsl{\,\raise.15ex\hbox{/}\mkern-.13.5mu D}
%
%
%
\font\numbers=cmss12
\font\upright=cmu10 scaled\magstep1
\def\stroke{\vrule height8pt width0.4pt depth-0.1pt}
\def\topfleck{\vrule height8pt width0.5pt depth-5.9pt}
\def\botfleck{\vrule height2pt width0.5pt depth0.1pt}
\def\Zmath{\vcenter{\hbox{\numbers\rlap{\rlap{Z}\kern
0.8pt\topfleck}\kern
2.2pt
                   \rlap Z\kern 6pt\botfleck\kern 1pt}}}
\def\Qmath{\vcenter{\hbox{\upright\rlap{\rlap{Q}\kern
                   3.8pt\stroke}\phantom{Q}}}}
\def\Nmath{\vcenter{\hbox{\upright\rlap{I}\kern 1.7pt N}}}
\def\Cmath{\vcenter{\hbox{\upright\rlap{\rlap{C}\kern
                   3.8pt\stroke}\phantom{C}}}}
\def\Rmath{\vcenter{\hbox{\upright\rlap{I}\kern 1.7pt R}}}
\def\Z{\ifmmode\Zmath\else$\Zmath$\fi}
\def\Q{\ifmmode\Qmath\else$\Qmath$\fi}
\def\N{\ifmmode\Nmath\else$\Nmath$\fi}
\def\C{\ifmmode\Cmath\else$\Cmath$\fi}
\def\R{\ifmmode\Rmath\else$\Rmath$\fi}

\Title{CLNS 95/1336, hep-th/9505086}
{\vbox{\centerline{Quantum Theory of Self-Induced Transparency  }}}
\bigskip
\bigskip

\centerline{Andr\'e LeClair}
\medskip\centerline{Newman Laboratory}
\centerline{Cornell University}
\centerline{Ithaca, NY  14853}
\bigskip\bigskip

\vskip .3in

It is known that classical electromagnetic radiation
at a frequency in resonance with energy splittings of
atoms in a dielectric medium can be described using the
classical sine-Gordon theory.   In this paper we quantize
the electromagnetic field and compute some quantum effects
by using known results from the
sine-Gordon quantum field theory.
In particular, we compute the intensity of spontaneously
emitted radition using the thermodynamic Bethe ansatz with
boundary interactions.

\Date{10/94}
%
%
%
%
%
%
\noblackbox

%
%
%
%
%
%
%
%
%
%

\newsec{Introduction}

The importance of integrable non-linear partial differential
equations in classical non-linear optics has been recognized for
some time.  The most well-known example concerns weakly non-linear
dielectric media, where only the first non-linear susceptibility
is considered important. In this situation, the envelope of the electric
field satisfies the non-linear Schrodinger equation.  This was
first understood theoretically by Hasegawa and Tappert\ref\rhas{A. Hasegawa
and F. Tappert, Appl. Phys. Lett. 23 (1973) 142.}.
The solitons predicted in \rhas\ were observed experimentally
in \ref\rmol{L. F. Mollenauer, R. H. Stollen and J. P. Gordon,
Phys. Rev. Lett. 45 (1980) 1095.}.
The occurence of these classical solitons in common optical fibers
promises to revolutionize high-speed telecommunications.

Our interest in this subject concerns the possibly interesting
quantum effects which arise when the electromagnetic field is
quantized.  The latter quantization amounts to studying
an interacting quantum field theory in the classical variables
which satisfy the non-linear differential equation.  These
quantum integrable models have been studied extensively,
and many of their properties have already been exactly computed.
A priori, one expects quantum effects to be small for macroscopically
large objects such as solitons\foot{In fiber optic systems the
soliton consists of a cluster of $10^8$ or more photons.}.
Nevertheless, using known exact results from the quantum non-linear
Schrodinger theory, quantum effects have been predicted and
measured\ref\rnla{S. J. Carter, P. D. Drummund, M. D. Reid and
and P. D. Drummond, Phys. Rev. Lett. 58 (1987) 1841.}\ref\rnlb{P. D.
Drummond and S. J. Carter, J. Opt. Soc. Am. B4 (1987) 1565.}\ref\rnlc{Y.
Lai
and H. A. Haus, Phys. Rev. A40 (1989) 844; H. A. Haus and Y. Lai, J. Opt.
Soc. Am. B7 (1990) 386.}\ref\rnld{M. Rosenbluh and R. M. Shelby,
Phys. Rev. Lett. 66 (1991) 153.}.

It is well-known that
the non-linear dielectric susceptibilities are enhanced when the radiation
is in resonance with the energy splitting of quantum states of the
atoms of the sample.
Near resonance, one is no longer in the weakly non-linear regime
(higher susceptibilities involving higher powers of the electric
field become as important) and the physics is no longer well described
by the non-linear Schrodinger equation.  Remarkably, as was understood
and demonstrated experimentally by McCall and Hahn\ref\trans{S. McCall
and E. L. Hahn, Phys. Rev. 183 (1969) 457.}, the system is well
described classically by another famous integrable equation,
the sine-Gordon (SG) equation.  The phenomenon is referred to as
`self-induced transparency'.

In this paper, we study the quantum effects which arise when
electric fields in resonance with a dielectric medium are quantized
by using known exact results for the SG quantum field theory.
In particular, we compute the intensity of spontaneously emitted
radiation by using the thermodynamic Bethe ansatz with
boundary interactions\foot{The first few sections of this paper
previously appeared in \ref\old{A. LeClair, {\it Quantum
Solitons in Non-Linear Optics: Resonant Dielectric Media},
CLNS 94/1302, hep-th/9410037, to appear in
{\it Unified Symmetry: In the Small and in the Large II},
Coral Gables 1995 conference, B. N. Kursunoglu, S. Mintz and A.
Perlmutter, eds.,
Plenum Publishing.}.}.  A different but related model is the
Dicke Model, which was solved using Bethe ansatz in \ref\rrup{V.
I. Rupasov and V. I. Yudson, Sov. Phys. JETP 60 (1984) 927.}.

\newsec{Classical Theory}

\def\sig{\sigma}
\def\cb{\bar{c}}
\def\bcl{\beta_{cl}}
\def\ce{\CE}
\def\om{\omega}
\def\svev#1{\vev{\sig_{#1}}}

In this section we review the manner in which the classical
sine-Gordon equation arises in resonant dielectric media\trans.

We consider electromagnetic radiation of frequency $\omega$
propagating through a collection of atoms, where the frequency
$\om$ is in resonance with an energy splitting of the atomic
states.  For simplicity, we suppose each atom is a two state
system described by the hamiltonian $H_0$ with
the following eigenstates: $H_0 \ket{\psi_1} = -\inv{2} \hbar
\om_0 \ket{\psi_1}$,
 $H_0 \ket{\psi_2} = \inv{2} \hbar
\om_0 \ket{\psi_2}$, such that $\hbar \om_0$ is the energy difference
of the two states.  In the presence of radiation, the atomic
hamiltonian is
\eqn\eIIi{H_{atom} = H_0 - \vec{p} \cdot \vec{E} , }
where $\vec{p} = e \sum_i \vec{r}_i $ is the electric dipole
moment operator.

We assume the radiation is propagating in the $\hat{x}$ direction,
and $\vec{E} = \hat{n} E(x,t)$, where $ \hat{n} \cdot \hat{x} = 0$.
The only non-zero matrix elements of the operator $\vec{p}\cdot \vec{E}$
can be parameterized as follows:
\eqn\eIIib{
\bra{\psi_2} \vec{p} \cdot \vec{E} \ket{\psi_1}  = p E(x,t) e^{-i\al} ,}
where
$p$ and $\al$ are constants which depend on the atom in question.
(We have assumed spherical symmetry.)  It is convenient to introduce
the Pauli matrices $\sig_i$, and write the hamiltonian as
\eqn\eIIii{
H_{atom} = - \inv{2} \hbar \om_0 \, \sig_3 - E(x,t) (p_1 \sig_1 + p_2
\sig_2 ) , }
where
$p_1 = p \cos \al , ~ p_2 = p\sin \al $.

The dynamics of the system is determined by Maxwell's equations,
\eqn\eIIiii{
\( \d_x^2 - \inv{\cb^2} \d_t^2 \) E(x,t) = \frac{4\pi}{c^2} ~
\d_t^2 P(x,t) ,}
where
$\cb^2 = c^2 / \ep_0$, $\ep_0$ is the ambient dielectric constant,
and $\vec{P} = \hat{n} P(x,t)$ is the dipole moment per unit volume.
The latter polarization can be expressed in terms of the expectations
of the Pauli spin matrices $\vev{\sig_i} = \bra{\psi} \sig_i \ket{\psi}$,
where $\ket{\psi}$ is the atomic wavefunction.  Namely,
\eqn\eIIiv{
\vec{P} = \hat{n}  n \( p_1 \vev{\sig_1} + p_2 \vev{\sig_2} \),}
where
$n$ is the number of atoms per unit volume\foot{To be more precise,
$\vev{\sig_i}$  here represents
average over many atoms in a small  volume, and is thus
a continuous field depending on $x,t$.}.      Thus, in addition
to the Maxwell equation \eIIiii, one has dynamical equations for
the polarization $P(x,t)$ which are determined by Schrodinger's
equation for the atom:
\eqn\eIIv{
i\hbar \,  \d_t \vev{\sig_i} = \bra{\psi} \[ \sig_i , H_{atom} \] \ket{\psi} .}
The latter can be expressed as
\eqn\eIIvi{
\d_t \vev{\sig_i} = \sum_{j,k} \varepsilon_{ijk}  V_j ~ \vev{\sig_k} , }
where
$\varepsilon$ is the completely antisymmetric tensor with
$\varepsilon_{123} = 1$, and
\eqn\eIIvii{
V_1 = \frac{ 2 E(x,t)}{\hbar} p_1 , ~~~~~
V_2 = \frac{ 2 E(x,t)}{\hbar} p_2 , ~~~~~
V_3 = \om_0 .}

To summarize, the dynamics is determined from the coupled equations
of motion \eIIiii\ and \eIIvi, wherein the atoms are treated
quantum mechanically and the radiation is classical.

\def\spar{\svev{\parallel}}
\def\sper{\svev{\perp}}

Let \eqn\Eeq{
E(x,t) = \ce (x,t) \cos (\om t - k x) }
where
$\om / k = \cb$ and $\ce (x,t)$ is the envelope of the electric
field.  We will assume the envelope is slowly varying in comparison
to the harmonic oscillations: $\d_t \ce \ll \om \ce, ~
\d_x \ce \ll k \ce$.  In this approximation, one finds
\eqn\eIIviii{
\( \d_x^2 - \inv{\cb^2} \d_t^2 \) E(x,t) \approx
\frac{ 2\om}{\cb}
\[ \(\d_x + \inv{\cb} \d_t \) \ce (x,t) \] \sin (\om t - kx ) .}

\def\ar{( \om t - k x + \al) }
Let us define $\spar, \sper$ as follows
\eqn\eIIix{\eqalign{
\spar &= \svev{1} \cos \ar + \svev{2} \sin \ar \cr
\sper &= - \svev{1} \sin \ar + \svev{2} \cos \ar . \cr
}}
We make the further approximation that
$\cos 2\ar  $ terms in the equations of motion
for $\d_t \svev{i}$ can be dropped in comparison to
$\cos \ar$ (and similarly for the sine terms)\foot{It can
be shown in perturbation theory that this is a good approximation
at or near resonance.}.  (This amounts to replacing
$\sin^2 \ar , \cos^2 \ar $ by $1/2$).  One then finds
\eqna\eIIx
$$\eqalignno{
\d_t \spar &= (\om - \om_0 ) \,  \sper  &\eIIx {a} \cr
\d_t \sper &= - \frac{\ce (x,t)}{\hbar} p \, \svev{3}
+ (\om_0 - \om ) \spar  &\eIIx {b} \cr
\d_t \svev{3}  &=  \frac{\ce (x,t)}{\hbar} p\,  \sper  . &\eIIx {c} \cr
}$$
Finally, equations \eIIiii\ and \eIIviii, upon making an approximation
analagous to the slowly varying envelope on the RHS of \eIIiii,
lead to
\eqn\eIIxi{
\[ (\d_x + \inv{\cb} \d_t ) \ce (x,t) \]
\sin (\om t - k x )
= \frac{2\pi }{c^2} \cb n \om p
\( \sin (\om t - kx) \sper - \cos (\om t - kx ) \spar \) . }

In order to solve these equations, note that
$\d_t (\sum_i \svev{i} \svev{i} ) = 0$.  Thus, if the atoms start
out in their ground state $| \psi_1 \rangle $,
with $\svev{3} = 1$, one has
$\sum_i \svev{i}^2 = 1$ for all time.  On resonance, when
$\om = \om_0$, eq. \eIIx{a} implies $\spar = 0$ for all time.
The constraint $\sum_i \svev{i}^2  = 1$
can be imposed with the following parameterization:
\eqn\eIIxii{ \sper =  \sin (\bcl \phi (x,t) ) , ~~~~~
\svev{3} =  \cos (\bcl \phi (x,t) ).}
The parameter $\bcl$ is arbitrary at this stage, but will be fixed
in the next section.
Equations \eIIx{b,c}\   now imply
\eqn\eIIxiii{
\d_t \phi = - \frac{ p}{\bcl \hbar} ~ \ce (x,t) .}
Inserting \eIIxiii\ into \eIIxi\ and defining
\eqn\frame{x'=2x - \cb t,}
one obtains the SG equation:
\eqn\eIIxiv{
\( \d_t^2 - \cb^2 \d_{x'}^2 \) \phi = - \frac{\mu^2}{\bcl}
\sin (\bcl \phi ) , }
where
\eqn\eIIxv{
\mu^2 = \frac{2\pi n p^2 \om}{\hbar \ep_0} . }

\newsec{The Quantum Spectrum}

In this section we proceed to quantize the electromagnetic field.
In order to do this, the SG field $\phi$ must be properly normalized
such that the energy of soliton solutions corresponds to the
physical energy;  this amounts to properly fixing the constant
$\bcl$.

The action which gives the classical SG equation of motion is
\eqn\eIIIi{
S_{SG} = \inv{\cb} \int dx' dt \(
\inv{2} (\d_t \phi)^2 - \frac{\cb^2}{2} (\d_{x'} \phi )^2
+ \frac{\mu^2}{\bcl^2} ~ \cos (\bcl \phi ) \) . }
On the other hand, the properly normalized Maxwell action is
\eqn\eIIIii{
S_{\rm Maxwell} = \inv{c^2} \int d^3 x dt ~ \( - \inv{4}
F_{\mu\nu} F^{\mu\nu} + ... \)  ~~~=
\inv{c^2} \int d^3 x dt ~ \inv{2} (\d_t A )^2 + .... }
where the vector potential is $\vec{A} = \hat{n} A $ and
as usual $\vec{E} = -\inv{c} \d_t \vec{A} $.  (One can show
that $A_0$ can be set to zero.)  One should be able to show directly
that the complete Maxwell action, upon making the appropriate
approximations of the last section, reduces to the sine-Gordon action,
since we have already shown this to be the case at the level
of the equations of motion.  Assuming this to be the case, in
order to properly normalize the field
$\phi$, one needs only compare the kinetic terms in
\eIIIi\ and \eIIIii.  The dimensional reduction is made by
assuming simply that $A$ is independent of $y,z$ and
$\int dydz = \CA$, where $\CA$ is an effective cross-sectional
area perpendicular to the direction of propagation (as in the
cross-sectional area of a fiber).  From \eIIxiii\ one has
\eqn\eIIIiii{
S_{\rm Maxwell} = \CA \( \frac{\bcl \hbar}{p} \)^2
\int dx dt ~  \inv{2} (\d_t \phi )^2  \cos^2 (\omega t - kx ) + .... }
In the above equation, we make an approximation analagous to the one made
in arriving at \eIIx{}   and replace $\cos^2 (\omega t - kx)$ by
$1/2$.  Noting that $dx dt = dx' dt /2 $,
upon comparing with \eIIIi, one fixes $\bcl^2 = 4p^2 /(\CA \hbar^2 \cb)$.

Finally, as is conventionally done in the quantum SG literature,
we rescale $\phi \to
\sqrt{\hbar} \phi$, so that $S_{SG} / \hbar$ takes the form \eIIIi
with $\bcl$ replaced by $\beta$, where
\eqn\eIIIvi{
\beta^2 = \hbar \bcl^2 = \frac{4 p^2 \sqrt{\ep_0}}{\CA \hbar c} . }
The constant $\beta$ is the conventional dimensionless coupling
constant in the quantum SG theory, and is allowed to be in
the range $0\leq \beta^2 \leq 8\pi$, where
$\beta^2 = 8\pi$ corresponds to a  phase
transition\ref\rcole{S. Coleman, Phys. Rev D11 (1975) 2088.}.
The limit where the radiation
is classical but the atom is still quantum mechanical corresponds
to the limit $\beta \to 0$.

\def\bep{\frac{\beta^2}{8\pi}}

The classical soliton solutions to the SG equation are characterized
by a topological charge $T= \pm 1$, where
$$T= \frac{\beta}{2\pi }  (\phi (x' = \infty) - \phi (x' = -\infty ) ) .$$
Solitons
of either charge correspond to solutions where at fixed $x$ the atoms in
the far past are in their ground state, and are all in their excited
state at some intermediate time:  $\svev{3}_{t= \pm \infty} = 1$.
These classical solitons have been observed experimentally in \trans.
What distinguishes solitons ($T=1$) from antisolitons ($T=-1 $)
is the sign of the envelope of the electric field.  From the known
classical soliton solutions and \eIIxiii\ one finds for
$T= \pm 1$,
\eqn\eIIIviii{
\ce (x' ,t) = \pm \frac{2 \hbar \mu}{p} \sqrt{ \frac{\cb + v}{\cb -v} }
\(  \cosh \(  \frac{\mu (x' -vt)}{\sqrt{\cb^2 - v^2}} \) \)^{-1}  .  }
Thus the electric fields for the soliton versus the antisoliton
are out of phase by $\pi$.

The particle spectrum of the quantum field theory is known\foot{The
literature on the quantum SG theory is extensive.  See for example
\ref\rzz{A. B. Zamolodchikov
and Al. B. Zamolodchikov, Ann. Phys. 120 (1979) 253.}, and references
therein.}
to consist of soliton and anti-soliton of mass $m_s$, and a
spectrum of breathers of mass
\eqn\ebreath{
m_n = 2 m_s ~ \sin \frac{n\gamma}{16} , ~~~~~~n= 1, 2, ....< \frac{8\pi}
{\gamma} , }
where
\eqn\egamma{
\gamma = \frac{\be^2}{1-\bep} .}
The n-th breather disappears from the spectrum when
$8\pi /\gamma = n$;  at this threshold the mass of this breather
is twice the soliton mass.  The n-th breather can thus be
considered as a bound state of n soliton/antisoliton pairs.
The $n=1$ breather is known to correspond to the particle associated
with the sine-Gordon field $\phi$ itself.

 From  \eIIIii\eIIIiii, one sees that the sine-Gordon scalar field
is essentially the photon vector potential, thus the $n=1$
breather may be thought of as a `envelope photon', which is
massive due to the dielectric properties of the medium.
Of course, this photon has little to do with the real free
photons of energy $\hbar \omega$.
The n-th breather state can then be viewed as an
n-envelope-photon bound state.  Under this heuristic assignment of
photon number to the spectrum, the soliton has photon number
$1/2$.  In specific physical situations, $\beta$ is very
small (see below), thus  $m_1 << m_s$, which indicates the
strong binding energy of a pair of solitons to form an envelope
photon.

The soliton mass is directly related to the parameter $\mu$ in
\eIIxv.
In the quantum theory, short distance singularities
are removed by suitably normal ordering the $\cos (\beta \phi )$
potential\rcole:
$$\frac{\mu^2}{\beta^2}  \cos (\beta \phi )  \to \lambda
: \cos (\beta \phi ) :$$  The anomalous scaling dimension
of the operator $:\cos (\beta \phi ) :$ is $\beta^2 / 4 \pi $,
so that $\lambda$ has mass dimension $2 (1- \bep )$.
Therefore,
\eqn\ems{m_s  \propto (\mu) ^{1/(1-\bep)} . }
This allows us to obtain quantum corrections to the frequency
and density dependence of the soliton mass.
Since $\mu \propto
\sqrt{\om n}$, one finds
\eqn\eIIIxi{
m_s  \propto
\sqrt{\om n} \( 1 + \frac{\beta^2}{16\pi} \log (\om n )
+ O(\beta^4 ) \). }
The exact relation between $m_s$ and $\mu$ is known\ref\ralzam{Al.
Al. B. Zamolodchikov, {\it Mass Scale in
Sine-Gordon and its Reductions}, Montpellier preprint
LPM-93-06, 1993.}:
\eqn\eexc{
m_s = c(\be) m^{\gamma / \be^2 }  \Lambda^{-\gamma /8\pi } ,
{}~~~~~~ m =  \frac{\mu \hbar}{\bar{c}^2} , }
where $\Lambda$ is an ultraviolet cutoff, and
\eqn\ecbet{
c(\beta) =
\frac{ 2 \Gamma (\gamma /16\pi ) }{\sqrt{\pi} \Gamma (\gamma /2\be^2 ) }
\[  \frac{ \pi}{2\be^2}
\frac{\Gamma ( 1- \bep )}{\Gamma ( \gamma^2 / 8\pi \be^2 ) }
\]^{ \gamma/2 \be^2 }  . }
As $\beta \to 0$, one obtains the well known classical expression,
\eqn\eclass{
m_s  =  \frac{8 m}{\beta^2}  ,      ~~~~~~(\beta \to 0 ). }
 From \ebreath, one sees that
for small $\beta$, $m$ is approximately the mass of the lowest breather.

The magnitude of the quantum corrections to classical results is
determined by the parameter $\beta^2/8\pi$.
In the original experiment of McCall and Hahn, a ruby sample was
used with $p= 4.8 \times 10^{-21}$ in cgs units, which corresponds
to $\beta^2 = 10^{-23} \sqrt{\ep_0} /\CA $.  The ruby rod had
a cross-sectional area of $\CA \approx 1 cm^2$.
Taking the density $n$ to be that of the $Cr^{+3}$ doping atoms
in the ruby, then $n \approx 10^{19} /cm^3$.  For
$\omega \approx 10^{15} s^{-1}$, one finds
$m \approx 10^{-37} g$, and $m_s \approx 10^{-13} g $.
This large hierarchy of mass scales between the soliton and
the lowest breather is due to the smallness of $\beta$.
Thus the experiment performed by McCall and Hahn is in
the extreme quasi-classical limit, with an extensive spectrum
of light and heavy breathers and
relatively heavy solitons with macroscopic
masses.  In the quantum theory, the classical soliton observed
in for example \trans, becomes a fundamental particle, i.e.
it is not to be thought of as a coherent state of photons,
due to the fact that the soliton carries non-zero conserved
topological charge.

Since in the limit $\beta \to 0$,  $m_s = 8m/ \beta^2$,
from \eIIIvi\ one can express
$\beta^2 /8\pi$ in terms of the classical soliton mass:
\eqn\eIIIix{
\bep = 16 \sqrt{\ep_0} \( \frac {\hbar \om}{m_s c^2} \)
\( \frac{\hbar n \CA c}{m_s c^2}\) . }
Thus,
\eqn\eIIIx{
\bep \sim ~~  \inv{N_\gamma} \( \frac{\lambda_c}{L_{\rm atom}} \),
}
where $N_\gamma = \hbar \om / m_s c^2$ roughly corresponds classically
to the number of photons that comprise the soliton,
$\lambda_c$ is the Compton wavelength of the soliton, and
$L_{\rm atom} = 1/n\CA$ is the inter-atomic spacing.  The above
equation summarizes where one expects quantum effects to be
important: when the soliton is comprised of small numbers of photons,
or when the Compton wavelength is large compared to the space
between the atoms.

The classical scattering matrix for the solitons has been computed
by comparing N-soliton solutions in the far past and far future\ref\rkor{V.
E. Korepin and L. D. Faddeev, Theor. Mat. Fiz 25 (1975) 147.}.
The exact quantum S-matrix is also known\rzz.
Since
$\beta^2 / 8\pi $ is small here,
the quantum corrections to classical scattering
are most easily determined by incorporating the one-loop corrections
to the classical scattering, which amounts to the
replacement $\beta^2 \to \gamma $. The
necessary formulas can be found in the above papers.

The most interesting aspect of the quantum scattering of solitons
is
that the so-called reflection amplitude is a purely quantum effect,
analagous to barrier penetration \ref\rkorb{V. E. Korepin,
Teor. Mat. Fiz. 34 (1978) 3.}.
Namely, one considers an in-state consisting of a soliton
of momentum $p_1$ and an antisoliton of momentum $p_2$ which scatters
into an out state where the momenta $p_1$ and $p_2$ are interchanged.
The $T= +1$
soliton is thus reflected back with momentum equal to that of
the incoming $T=-1$ antisoliton.  In the semi-classical approximation,
this reflection amplitude is
\eqn\eIIIxii{
S_R (\theta ) = \inv{2} \( e^{16\pi^2 i /\gamma } - 1 \)
e^{- \frac{8\pi }{\gamma} |\theta | }  ~ S(\theta )
}
where
\eqn\eIIIxiii{
S(\theta ) = \exp \( \frac{8}{\gamma} \int_0^\pi
d\eta
\log \[ \frac{e^{\theta - i \eta} + 1}{e^\theta + e^{-i\eta} } \]
\), }
and
$\theta = \theta_1 - \theta_2 $, where
$p_{1,2} = M_s \sinh \theta_{1,2} $.
Since $1/\gamma \approx 1/\beta^2$ is very large, the oscillatory
factor $\exp( 16\pi^2 i/\gamma ) - 1$ in \eIIIxii\
will make the detection of reflection processes difficult,
since even in
a small range of $\beta^2$, this factor averages to zero.

\newsec{Spontaneous Emission}

In this section we will consider the spontaneous emission of
radiation from a collection of atoms in their excited state.
We suppose that all the atoms have been `pumped' by some means
to their excited state at $t=0$.  In the coordinate system
$(x' , t' )$, defined by
\frame\ and $t'=t$, the atoms are along the $x'$ axis at
$t=0$.  If $|\psi \rangle$ is the initial state corresponding
to all atoms being in their excited state, then
\eqn\eIVi{
\langle \psi | \sigma_3 | \psi \rangle = -1 = \cos \beta \phi
, ~~~~~~~(t=0).}
Thus, the initial state corresponds to the initial constant
field configuration
\eqn\eIVii{
\phi = \phi_0 =  \frac{\pi}{\beta} .  ~~~~~~~(t=0). }
All atoms initially in the ground state corresponds to
$\phi_0 = 0$.  The subsequent time development of the
system thus amounts to an initial value problem in the
SG quantum field theory.

The initial condition \eIVii\ may be imposed by adding to the
SG action the boundary term:
\eqn\eIViii{
S_{\rm boundary} =  g \int dx \cos \( \beta (\phi - \phi_0 ) /2 \). }
In the limit $g\to \infinity$, this enforces the boundary condition
$\phi = \phi_0$.
It is known that the boundary interaction \eIViii\ preserves the
integrability of the SG theory\ref\rgz{S. Ghoshal and A. B.
Zamolodchikov, Int. J. Mod. Phys. A9 (1994) 3841.}.
Ghoshal and Zamolodchikov have developed a comprehensive approach
to studying integrable quantum field theory with boundary interactions,
and in particular have solved the boundary SG theory.  The limit
$g\to \infinity$ is referred to as the fixed boundary condition
in \rgz, whereas $g=0$ is called the free boundary condition.
As shown there, information about the boundary condition
is concentrated in a boundary state $|B\rangle$ at $t=0$,
with the general form
\eqn\eIViv{
|B\rangle  = \exp \(
\int d\th K^{ab} (\th ) A_a (-\th ) A_b (\th ) \)
 |0 \rangle }
where $A_a$ are creation operators of particles of type $a$ with
rapidity $\th$, and
\eqn\eIVivb{
K^{ab} (\th ) = R^b_{\bar{a}} (i\pi /2 - \th ) ,}
where  $R(\th )$ are the S-matrices for reflection off
the boundary, and $\bar{a}$ is the charge conjugate of $a$.
The rapidity parameterizes energy and momentum as follows,
\eqn\eIVv{
E' = m \cosh \th , ~~~~~~~P' = m \sinh \th  , }
where
$E' , P'$ are energy and momentum in the coordinate system
$(x' , t' )$.  (We have set $\hbar = \bar{c} = 1$.)
In our application, since the boundary is a spacial axis at
$t=0$, $K^{ab} (\th )$ represents an amplitude for emission
of particle pairs at the boundary.

In order to compute the intensity, one should first study
a partition function with the relevant boundary conditions.  In
euclidean space, one may consider the partition functions
\eqn\epart{
Z(R)  =  \langle B' | e^{-H R} | B \rangle ,}
where $H$ is the hamiltonian, and $\langle B' |$ is a conjugate
boundary state.   Taking  $| B \rangle$ to be the boundary
state for the fixed boundary condition with $\phi_0$ given in
\eIVii, and $\langle B' |$ to be the boundary state for
the free boundary condition provides information on the
time evolution of a state where the atoms are all initially
excited,  and after a time interval $R$ the final state
has no constraints on $\phi$.

\def\ep{\varepsilon}

To simplify the initial discussion, let us assume there is
only one type of particle in the spectrum, with bulk
two-particle S-matrix $S(\th )$, and boundary S-matrices
$K (\th )$ and $K' (\th )$ for the boundary states
$|B \rangle$ and $|B' \rangle$ respectively.
The function $Z(R)$ has the interpretation as a partition function
on a strip of width $R$ with boundary conditions $|B\rangle$
at both ends.  As shown in \ref\rlec{A. LeClair, G. Mussardo, H.
Saleur and S. Skorik, {\it Boundary Energy and Boundary
States in Integrable Quantum Field Theories}, CLNS/95-1328,
USC-95-005, ISAS/EP/95-26, hep-th/9503227.}, if
one uses the expression \eIViv\ for $|B\rangle $, then one
encounters divergences in $Z(R)$ due to the fact that the
formula \eIViv\ is only valid in infinite volume.
We thus regulate these divergences by taking the boundary
at $t=0$ to have a large but finite length $L$.  ($L$ corresponds
to the length of the sample.)  Then $Z(R)$ is
characterized by the thermodynamic Bethe ansatz (TBA)\rlec.
These TBA equations generalize the bulk equations with
no boundaries
derived in \ref\ralz{Al. Zamolodchikov, Nucl. Phys. B342 (1990),
695.}.  One
introduces a density of pairs of particles per unit length
$\rho (\th )$ and similarly a  density of holes $\rho^h (\th )$.
Due to the finite volume $L$, the momenta are quantized in a way
that involves the bulk S-matrix:
\eqn\eIVx{
\rho (\th ) + \rho^h (\th ) = \frac{m}{2\pi } \cosh \th
+ \inv{2\pi}  \( \Phi * \rho \) (\th ),}
where
\eqn\eIVxi{
\Phi (\th ) = -i \d_\th \log S (\th ) , }
and $*$ denotes convolution,
\eqn\eIVxii{
\( \Phi * \rho \) (\th ) =
\int_{-\infinity}^\infinity d\th' \Phi (\th - \th') \rho (\th') . }
Defining
\eqn\eIVxiii{
\frac{\rho^h}{\rho} = e^\ep , }
then $\ep$
satisfies the integral equation
\eqn\eIVxiv{
\ep (\th ) = 2mR \cosh \th - \log ( K (K')^* ) -
\inv{2\pi} \Phi * \log \( 1 + e^{-\ep } \) . }
The partition function is then given by the
formula
\eqn\eIVxv{
\log Z (R) = \frac{mL}{4\pi} \int_{-\infinity}^\infinity
d\th \cosh \th \log \( 1 + e^{-\ep (\th )} \) .
}

One can now compute the intensity, i.e. the energy flux per unit
time of the emitted radiation.  This  radiation will consist
of waves traveling to the right and left.  Above, the SG theory
describes the envelope for the right moving wave
\Eeq, therefore we compute the intensity in the right-moving
radiation.  The TBA analysis in \rlec\ assumes periodic
boundary conditions in the spacial direction
(cylindrical geometry), however we consider the intensity computed
below to be a good model of the measureable intensity at one end
of the sample.

In the slowly varying envelope approximation, the Pointing vector
is
\eqn\eIVxxii{
\vec{S} = \frac{c \sqrt{\ep_0}}{4 \pi }
\CE^2 (x,t) \cos^2 (\omega t - kx ) \hat{x} . }
Averaging over periods of the $\cos \omega t$ oscillations,
one has that
\eqn\eIVxxiii{
I = \inv{2} I_{\rm env} , }
where
$I_{\rm env}$ is the intensity of the envelope.
$I_{\rm env}$ can be computed from the particle picture of
the above TBA.  Let $(E, P, v)$, $(E', P' , v')$ be the
energy, momentum and velocity of a single particle in the
$(x,t)$ and $(x', t')$ coordinate systems respectively.
One has
\eqn\eIVxxiv{
P = 2 P' , ~~~~~E= E' - \bar{c} P' , ~~~~~v = (v' + \bar{c} )/2 .}
If $E' , P'$ are given as in \eIVv, then
$E = me^{-\th } $, and $v = (1 + \tanh \th )/2 $.

In the coordinate system $(x' , t')$, pairs of particles
are emitted with rapidity $\th$ and $-\th$.  For both particles
in the pair, $v(\th ) >0 $, thus
\eqn\eIVxxv{\eqalign{
I_{\rm env} &= \int_0^\infinity d\th \rho (\th )
\( E(\th ) v(\th ) + E(-\th ) v(-\th ) \) \cr
&= \frac{m\bar{c}^3}{2}
\int_{-\infinity}^\infinity d\th  ~ \rho(\th ) \inv{ \cosh \th } . \cr}}
(We have reinstated $\bar{c}, \hbar$;  $\rho$ has units
of $({\rm length})^{-1} $ and satisfies the equation \eIVx\ with
$m \to m\bar{c} /\hbar $. )

In the SG case, for general $\beta$, the above TBA equations
are very complicated due to the non-diagonal pieces
(reflection pieces) of the soliton S-matrix.  To simplify
the situation, we recall that $\beta^2 / 8 \pi$ is very small
in physical applications, and is thus well approximated by
\eqn\eIVxxvi{
\frac{\beta^2}{8 \pi}    = \inv{N+1} , }
for some integer $N$.  Namely, $N$ is taken as the integer which
best approximates \eIVxxvi.  At the points \eIVxxvi,
the reflection amplitudes vanish and the scattering becomes
diagonal.  As explained above, for the problem of spontaneous
emission, one should take the boundary state $\langle B' |$ to
be for the free boundary condition.  This leads to the problem
that the solitons, though they scatter diagonally off the
fixed boundary condition,  scatter off-diagonally from the
free boundary condition, and it is not known how to formulate
TBA equations in this situation.  Instead we solve a different problem
corresponding to an initial state wherein all the atoms are in
their excited state, and a final state where all the atoms are in
their ground state.  Both of these are fixed boundary conditions
with diagonal soliton scattering off the boundary.

The TBA equations become the following.  Introduce
densities of particles and holes per unit length for each
type of particle $\rho_a (\th ) , \rho^h_a (\th ) $ where
$a$ runs over breather and soliton indices:
$a\in \{ 1, 2, ..., N-1, + , - \} $, where $\pm$ refers to the
soliton and anti-soliton.  Introducing
\eqn\eIVxxvii{
\frac{\rho_a^h}{\rho_a}  = e^{\ep_a } , }
one has
\eqn\eIVxxviii{
\rho_a + \rho_a^h = \frac{m_a}{2\pi} \cosh \th
+ \inv{2\pi} \Phi_{ab} * \rho_b }
\eqn\eIVxxix{
\ep_a = \nu_a - \inv{2\pi} \sum_b \Phi_{ab} * \log ( 1 + e^{-\ep_b} ) }
\eqn\eIVxxx{
\log Z(R) = \frac{L}{4\pi}  \sum_a
\int_{-\infinity}^\infinity  d\th ~ m_a \cosh \th
\log (1 + e^{-\ep_a} ) }
\eqn\eIVxxxi{
I_{\rm env} = \sum_a  \frac{m_a \bar{c}^3 }{2}
\int_{-\infinity}^\infinity d\th \rho_a (\th ) \inv{\cosh \th } , }
where
\eqn\eIVxxxii{
\nu_a  = 2 m_a R \cosh \th - \log (K_a (K'_a)^* ) , }
\eqn\eIVxxxiii{
\Phi_{ab} = -i \d_\th \log S_{ab} (\th ) , }
where $S_{ab}$ is the 2-body S-matrix for particles of type
$a,b$ and $K_a$ are the boundary S-matrices for fixed boundary
condition $\phi_0 = \pi / \beta $, and $K'_a$ are those
for the fixed boundary condition $\phi_0 = 0$.

The SG scattering theory at these reflectionless points
is equivalent to the minimal scattering theory associated to the
root system of the $\hat{D}_{N+1}$ affine Lie algebra\ref\rcor{H. W.
Braden, E. Corrigan, P. E. Dorey, and R. Sasaki, Phys. Lett. B227 (1989) 411.}
\ref\rtim{T. R. Klassen and E. Melzer, Nucl. Phys. B338 (1990) 485.}.
There is a remarkable universal form of the above TBA equations\ref\ralzb{Al.
Zamolodchikov, Phys. Lett. B253 (1991) 391.}\ref\rdynk{F. Ravanini,
R. Tateo and A. Valleriani, Int. J. Mod. Phys. A8 (1993) 1707. }:
\eqn\eIVxxxiv{
\ep_a = \nu_a - \inv{2\pi} \sum_b N_{ab} \Phi * \( \nu_b -
\log \( 1 + e^{\ep_b} \) \) , }
where $\Phi$ is the universal kernel:
\eqn\eIVxxxv{
\Phi (\th ) = \frac{N}{\cosh N\th } , }
and $N_{ab}$ is the incidence matrix for the Dynkin diagram below.
Using similar manipulations as in \ralzb\rdynk, one can obtain
the universal form of the integral equation for
$\rho_a , \rho^h_a$:
\eqn\eIVxxxvi{
\rho_a + \rho_a^h  = \nu_a' + \inv{2\pi} \sum_b N_{ab}
\Phi * \( \rho_b^h  - \nu_b' \) , }
where
\eqn\eIVxxxvii{
\nu_a ' = \frac{m_a}{2\pi} \cosh \th .}

\midinsert
\epsfxsize = 3in
\bigskip\bigskip\bigskip\bigskip
\vbox{\vskip -.1in\hbox{\centerline{\epsffile{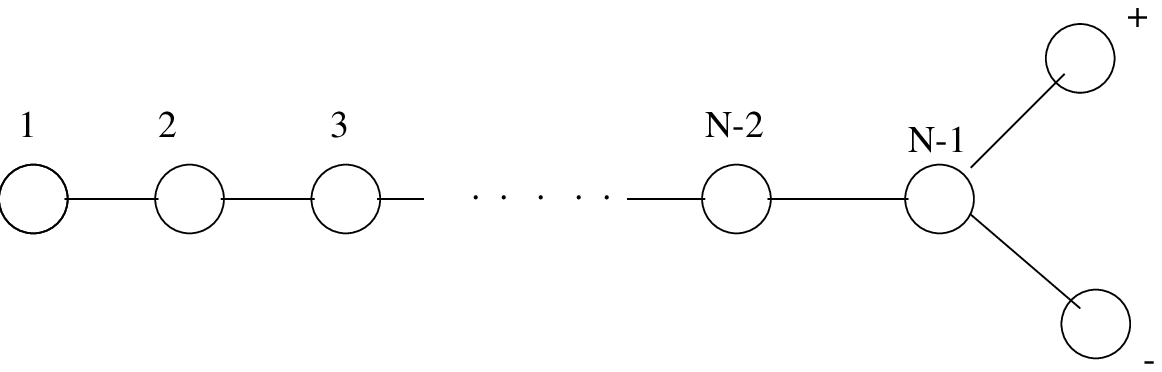}}}
\vskip .1in
{\leftskip .5in \rightskip .5in \noindent \ninerm \baselineskip=10pt
Figure 1.
Dynkin diagram for $\hat{D}_{N+1}$.
\smallskip}} \bigskip
\endinsert

The boundary S-matrices for the fixed boundary conditions
can be found in
\rgz\ref\rghosh{S. Ghoshal, Int. J. Mod. Phys. A9 (1994)
4801.}.  Specializing to the reflectionless
points, and using some Gamma function identities, one finds the
following\foot{The denominator of (5.23) in \rgz\ should
be $\Pi (x, \pi/2 ) \Pi (-x, \pi/2 ) \Pi (x, -\pi/2)
\Pi (-x , -\pi/2 ) $, and (5.24) should read
$\sigma(x, u) \sigma (x, -u ) = \cos^2 x
\[ \cos (x + \lambda u) \cos (x-\lambda u) \]^{-1} $.}.
We give below the $R$ reflection amplitudes;
the $K$ functions that appear in the integral equations above
are given by $K_a (\th ) = R_a (i\pi/2 - \th )$.
For the solitons, one has
\eqn\eR{
R_\pm (u) = (-1)^{N-1}
\frac{ \cos (\xi \pm Nu )}{\cos \xi }
\sigma (\xi , u )
\prod_{k=0}^{N-1}
\frac{ \sin ( \frac{u}{2} + \frac{\pi k}{4N} ) }{\sin ( \frac{u}{2}
- \frac{\pi k}{4N} ) } , }
where
$u= -i\th $, $\xi = \frac{4\pi}{\beta} \phi_0 $,
and
\eqn\esig{
\sigma (\xi , u ) = \prod_{k=0}^{N-1}
\frac{
\cos \( \frac{\xi}{2N} + \frac{ \pi (1+2k) }{4N}  \)
\cos \( \frac{\xi}{2N} - \frac{ \pi (1+2k) }{4N}  \)
}
{
\cos \( \frac{\xi}{2N} + \frac{ \pi (1+2k) }{4N} -u/2   \)
\cos \( \frac{\xi}{2N} - \frac{ \pi (1+2k) }{4N} +u/2   \)
} . }
For the n-th breather one has
\eqn\ebre{
R_n (u) = R^{(n)}_0 (u) R^{(n)}_1 (u) , }
\eqn\ebreb{\eqalign{
R_0^{(n)} (u) &= (-1)^{n+1}
\frac{
\cos \( \frac{u}{2} + \frac{n\pi}{4N} \)
\cos \( \frac{u}{2} - \frac{\pi}{4} - \frac{n\pi}{4N} \)
\sin \( \frac{u}{2} + \frac{\pi}{4} \) }
{\cos \( \frac{u}{2} - \frac{n\pi}{4N} \)
\cos \( \frac{u}{2} + \frac{\pi}{4} + \frac{n\pi}{4N} \)
\sin \( \frac{u}{2} - \frac{\pi}{4} \) }
\cr
& ~~~~~~~~~~~~~~\times
\prod_{k=1}^{n-1}
\frac{
\sin (u + \frac{k\pi}{2N} )
\cos^2 \(\frac{u}{2} - \frac{\pi}{4} - \frac{k\pi}{4N} \)
}{ \sin (u - \frac{k\pi}{2N} )
\cos^2 \(\frac{u}{2} + \frac{\pi}{4} + \frac{k\pi}{4N} \)
}
\cr }}
\eqn\ebrec{\eqalign{
R_1^{(2n)} (u) &=
\prod_{l=1}^n
\(
\frac{ \sin u - \cos \( \frac{\xi}{N} - (l-1/2) \frac{\pi}{N} \) }
{ \sin u + \cos \( \frac{\xi}{N} - (l-1/2) \frac{\pi}{N} \) }
\)
\(
\frac{ \sin u - \cos \( \frac{\xi}{N} + (l-1/2) \frac{\pi}{N} \) }
{ \sin u + \cos \( \frac{\xi}{N} + (l-1/2) \frac{\pi}{N} \) }
\)  \cr
&~~~~~~~~~~n = 1, 2, \cdots <N/2 \cr }}
\eqn\ebred{\eqalign{
R_1^{(2n-1)} (u) &=
\frac{ \cos (\xi /N ) - \sin u }
{ \cos (\xi /N ) + \sin u }
\prod_{l=1}^{n-1}
\frac{ \sin u  - \cos \( \frac{\xi}{N} - \frac{l\pi}{N} \) }
{ \sin u  + \cos \( \frac{\xi}{N} - \frac{l\pi}{N} \) }
\frac{ \sin u  - \cos \( \frac{\xi}{N} + \frac{l\pi}{N} \) }
{ \sin u  + \cos \( \frac{\xi}{N} + \frac{l\pi}{N} \) }
\cr
&~~~~~~~~~~~~~~~~n= 1,2, ..., < \frac{N+1}{2}  .
\cr }}

\newsec{Concluding Remarks}

We have shown how quantized radiation interacting with
two-level atoms leads to the quantum sine-Gordon theory
in the deep semi-classical regime.  As expected, this implies
quantum corrections are quite small.  We have not considered
here the kinds of quantum effects which were studied previously
for the non-linear Schrodinger case, which involve taking into
account coherent state initial conditions of the radiation
in real systems and  the squeezing of such states.
For the resonant situation considered here, the topological
charge of the sine-Gordon solitons presents difficulties for
constructing analagous coherent states of photons.

An interesting theoretical question which arises in this work
concerns the classical limit, $\beta \to 0$, of the thermodynamic
quantities in the SG theory.  As suggested above, one can approach
this limit by hopping along the refectionless points \eIVxxvi.
Then the problem becomes that of the $N$ goes to infinity limit
of the $\hat{D}_{N+1}$  TBA equations.  Since this corresponds
to a certain classical limit of the SG theory,  this leads us
to believe this limit is meaningful.  The interesting
limit from a physical point of view is to take  $\beta \to 0$,
keeping  $m_s$, which tends to $ 8\mu / \beta_{\rm cl}^2 \bar{c}^2 $,
fixed.  This is reminiscent of
the large $N$ studies of the ground state energy for the
$SU(N)$ sigma models (principal chiral model), where analytic
expressions were obtained in terms of Bessel functions\ref\rweig{
V. A. Fateev, V. A. Kazakov, and P. B. Wiegmann, Phys. Rev. Lett.
73 (1994) 1750; Nucl. Phys. B424 (1994) 505.}.
Another interesting question concerns how sensitive
the classical limit is to the way in which $\beta$ tends to zero.
I.e. is the limit different if one does not hop along the
reflectionless points?   We hope to address these issues in
a future publication.

\bigskip

\centerline{\bf Acknowledgements}

It is a pleasure to thank Peter Lepage,
Sergei Lukyanov, Henry Tye and   Paul Wiegmann  for useful discussions.
This work is supported by an Alfred P. Sloan Foundation fellowship,
and the National Science Foundation in part through the
National Young Investigator program.

\listrefs
\end